\documentclass{article}
\usepackage{spconf,amsmath,graphicx}
\usepackage{booktabs}
\usepackage{xcolor}
\usepackage{amsmath}
\usepackage{amssymb}
\usepackage{mathtools}
\usepackage{amsthm}
\usepackage{bm}
\usepackage{parskip}
 
\usepackage{subfigure} 

\newtheorem{theorem}{Theorem}[section]

\title{Evaluating Strong Idempotence of Image Codec}
\name{Qian Zhang$^{\star\dagger}$ \qquad Tongda Xu$^{\star 1}$\thanks{1. Qian Zhang and Tongda Xu have equal contribution.} \qquad Yanghao Li$^{\star}$ \qquad Yan Wang$^{\star 2}$\thanks{2. To whom correspondence should be addressed.}}
  
\address{$^{\star}$ Institute for AI Industry Research (AIR), Tsinghua University \qquad $^{\dagger}$ Xi'an Jiaotong University}

\begin{document}
%
\maketitle
\begin{abstract}
In this paper, we first propose the concept of strong idempotent codec based on idempotent codec. The idempotence of codec refers to the stability of codec to re-compression. Similarly, we define the strong idempotence of codec as the stability of codec to multiple quality re-compression, which is an important feature of codec in the context of cloud transcoding. We provide a detailed example of strong idempotent codec with known source distribution. Further, we formalize a testing protocol to evaluate the strong idempotence of image codec. And finally, we evaluate the strong idempotence of current image codecs, including traditional codec and recent neural codec. Experimental results show that current image codecs are not stable to multi-round re-compression with different qualities, even if they are close to idempotent.
\end{abstract}
\section{Introduction}
\label{sec:intro}
Recently, the idempotence property of image codec has been frequently revisited \cite{kim2020instability,li2022improving}. The idempotence of image codec refers to the codec's ability to preserve the quality of reconstruction after multi-round re-compression. More specifically, we denote the original image as $\bm{x}$, the encoder-decoder pair as $f(.)$, and the reconstructed image as $f(\bm{x})$. Then, we say the codec is idempotent if it satisfies:
\begin{align}
    f(\bm{x})=f(f(\bm{x})) \label{eq:ide}
\end{align}
As addressed by previous works \cite{kim2020instability,li2022improving}, the traditional codec standards such as JPEG \cite{wallace1992jpeg} has good idempotence, and neural image compression's \cite{lee2018context,balle2018variational,minnen2018joint} idempotence can also be improved.
However, in practice, the idempotent codec remains inadequate for fully preserving image quality in multi-round re-compression. For example, for a social network scenario, users might have different devices and network condition. Different users might acquire the image, re-compress it by their own device at different bitrates, and then re-distribute the image to other users. Therefore, an image might be re-compressed multiple times at different bitrates and quality factors. In that case, an idempotent image codec might not be able to ensure image quality during the different quality re-compression.

To address this problem, we propose the concept of strong idempotence. It is an extension of idempotence by taking the quality factor into consideration. It generalizes idempotence and is more suitable for practical communication scenario. We formally define a metric to evaluate the strong idempotence and provide a testing protocol. Finally, we evaluate the strong idempotence of current image codecs using our metric. Experimental results show that multi-round recompression with different quality factors is challenging even for traditional codec that satisfies idempotence.

\section{Related Works}
For traditional codec, the idempotence property has been discussed since 2000 \cite{vasudev2000image}. The current JPEG, JPEG2000 and MPEG standards are not idempotent \cite{joshi2000comparison}, but their idempotence can be improved through clipping and adapting of transform matrix \cite{erdem1994multi,horne1996study,sorial1997multigeneration,yi2004research}. On the other hand, JPEG XS \cite{descampe2017jpeg, richter2017multi} in 2017 explicitly required multi-generational robustness, which was defined as no noticeable quality degradation for up to 10 encode-decode cycles. For traditional video codec, \cite{zhu2010idempotent} proposed an idempotent h.264 encoder by clipping the compensation matrix.

For more recent learned image compression (LIC), \cite{kim2020instability} first builds a benchmark on the idempotence of LIC method and proposes a loss function to improve the robustness. Based on this work,  \cite{kim2022successive} later proposes two conditions for the compression method to achieve idempotence. \cite{li2022improving} uses straight quantization, reversibility loss function and channel relaxation methods to reduce generation loss. However, all the above works focused on multi-round compression with the same quality. The robustness with multi-quality re-compression is not visited.
\label{sec:rw}

\section{Strong Idempotence}
\subsection{Definition and Properties of Strong Idempotence}
We want to use strong idempotence to describe the stability of image codec to re-compression with different quality factors. Formally, we denote the original image as $\bm{x}$, the encoder-decoder pair with quality factor $q$ as $f(.,q)$ and the reconstruction image as $f(\bm{x},q)$. We note that $q$ is the quality factor instead of the quantization step-size, so a low $q$ means low quality. Then, we say a codec is strong indempotent if it satisfies:
\begin{gather}
    f(\bm{x},q_{min})=f(f(\bm{x},q_1),...,q_k),\notag\\
    \textrm{where } q_{min} = \min \{q_1,...,q_k\}.  \label{eq:ide}
\end{gather}
Eq.~\ref{eq:ide} indicates that after multi-round re-compression with varying quality factor $q$, the strong idempotent codec's reconstruction is the same as the reconstruction of single pass compression with the lowest quality factor. Furthermore, A strong indempotent codec has following properties:
\begin{enumerate}
    \item \textbf{A strong indempotent codec is indempotent.} It is obvious that strong indempotence implies indempotence by setting $q_1,...q_k$ to same value. 
    \item \textbf{The bitrate of strong indempotent codec is implicitly bounded.} Despite only limiting the reconstruction quality, the bitrate is also implicitly secured. Consider an indempotent codec $f(.)$, then the bitrate of compressing $f(x)$ remains the same during the compression process. As strong indempotent codec is also indempotent, we can always re-compress $f(\bm{x},q_{min})$ without changing the bitrate. This means that we can add extra re-compression with quality $q_{min}$ to the right hand side of Eq.~\ref{eq:ide} to obtain the same bitrate as compressing $f(\bm{x},q_{min})$. This property secures that the bitrate of compressing $\bm{x}$ sequentially with $q_1,...,q_k$ is the same as compressing $f(\bm{x},q_{min})$.
\end{enumerate}
\subsection{A Toy Example}
Now we give a toy example of strong idempotent codec for a better understanding of its property. Consider we have a 1 dimension source signal $x\sim\mathcal{U}(0,1)$, which satisfies unit uniform distribution. Then for a codec with 4 scalar quantized codeword points and rate $\log 4$, a possible choice is to select 4 codeword points $x_1=1/8,x_2=3/8,x_3=5/8,x_4=7/8$ uniformly in region $[0,1]$. And any value within $[x_i-1/8,x_i+1/8]$ is quantized into $x_i$. We assume that this codec has $3$ quality points: $\log 4$ is the highest rate with quality $q_3$, $\log 2$ is the middle rate with quality $q_2$ and $\log 1$ is the lowest rate with quality $q_1$. Then the problem becomes how to design the codeword of $q_2$, $q_1$ to make the codec strong idempotent.
\begin{figure}[h]
\centering
\includegraphics[width=\columnwidth]{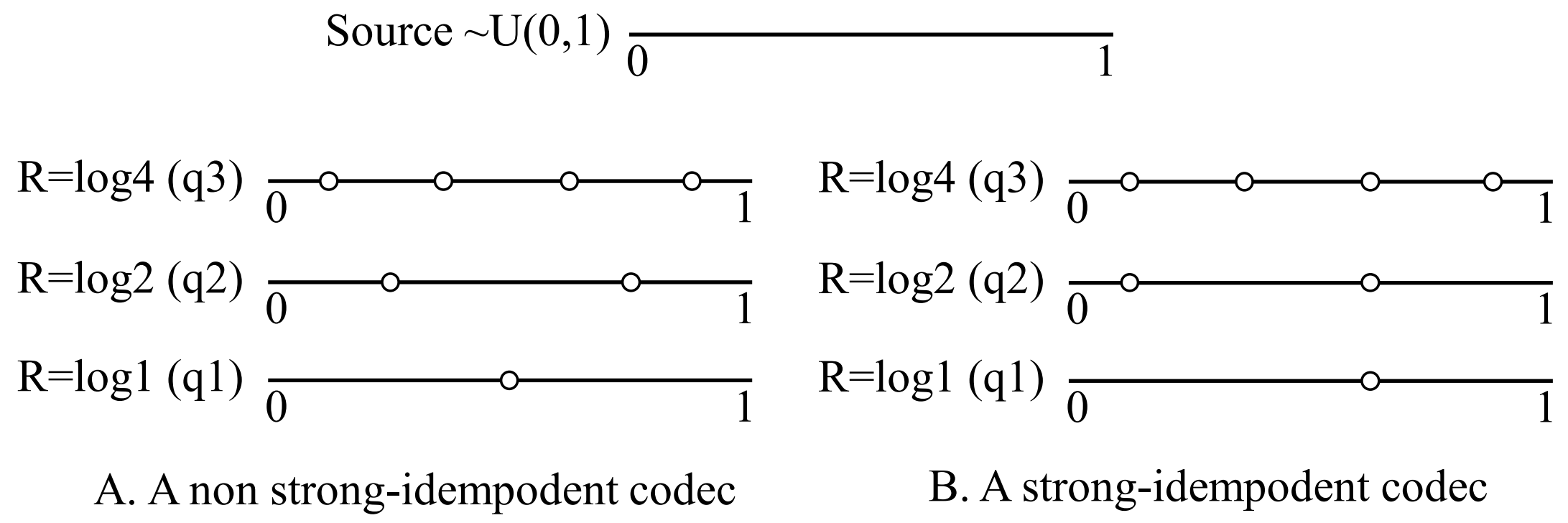}
\caption{A toy example of A. non-strong idempotent codec, B. a strong idempotent codec with the uniform source.}
\label{fig:01}
\end{figure}

Considering we design the codeword point with size 2 and 1 following size 4, we obtain the codec in Fig.~\ref{fig:01}.A. This codec is obviously non-strong idempotent. Assume that we compress $x$ with $q_1$ and then $q_3$, the final reconstruction will be either $3/8$ or $5/8$ instead of $1/2$. However, if we compress $x$ with $q_1$ alone, the reconstruction is $1/2$. This is because the codeword of $q_1$ is not a subset of codeword of $q_3$.

Consider an alternative design as Fig.~\ref{fig:01}.B. For $q_2$ with codeword number 2, we take a subset from $4$ codeword of $q_3$. Similarly, for $q_1$ with codeword number 1, we take a subset from $2$ codeword of $q_2$. And this codec is obviously strong idempotent. This is because for any two quality points $q_i,q_j$ with $i<j$, the codeword of $q_i$ is always a subset of the codeword of $q_j$. And the similar rule is true for vector-quantization.

\subsection{Designing a Metric}
Prior to the metric design, we first examine what will happen to the general non-idempotent codec when it is used for multi-round re-compressing. For general codec, the distortion of multi-round re-compressing with $\{q_1,...,q_k\}$ is not less than the distortion of single round compression with $q_{min}$. And for strong idempotence codec, those two distortions are equal to each other, and it is the best we can achieve. Formally, we have:
\begin{theorem}
\label{thm:dist}
The distortion of single round compression is not greater than the distortion of multi-round compression. In other words, denote $d(.,.)$ as a distortion metric, we have:
\begin{align}
    d(\bm{x},f(\bm{x},q_{min}))\le d(\bm{x},f(f(\bm{x},q_1),...,q_k)).
\end{align}
\begin{proof}
We proof by contradiction. First, let's assume that $d(\bm{x},f(\bm{x},q_{min}))> d(\bm{x},f(f(\bm{x},q_1),...,q_k))$ holds. And from data processing inequality, we have:
\begin{align}
    I(\bm{x},f(\bm{x},q_{min}))\ge I(\bm{x},f(f(\bm{x},q_1),...,q_k)),
\end{align}
which means that exist an achievable rate distortion pair with rate $\le I(\bm{x},f(\bm{x},q_{min}))$ and distortion $<d(\bm{x},f(\bm{x},q_{min}))$. And this is in contradiction to the converse of rate distortion theorem \cite{cover1999elements}. Therefore, the assumption is wrong, which completes the proof.
\end{proof}
\end{theorem} 

\begin{figure*}
\label{fig:02}
\begin{minipage}[t]{1.0\linewidth}
    \centering
    \subfigure[]{\label{fig:02a}\includegraphics[width=11.1cm,height=6.8cm]{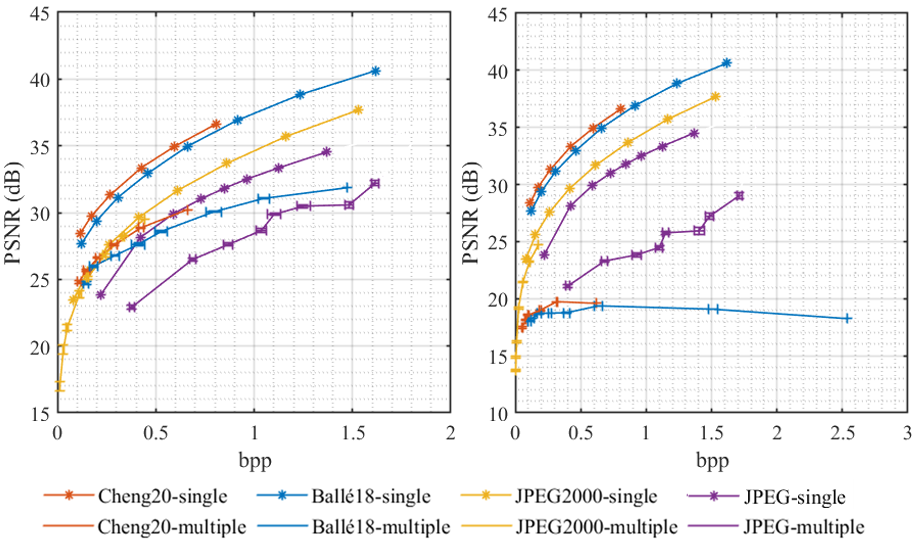}\hspace{0.1cm}}
    \subfigure[]{\label{fig:02b}\includegraphics[width=6.4cm]{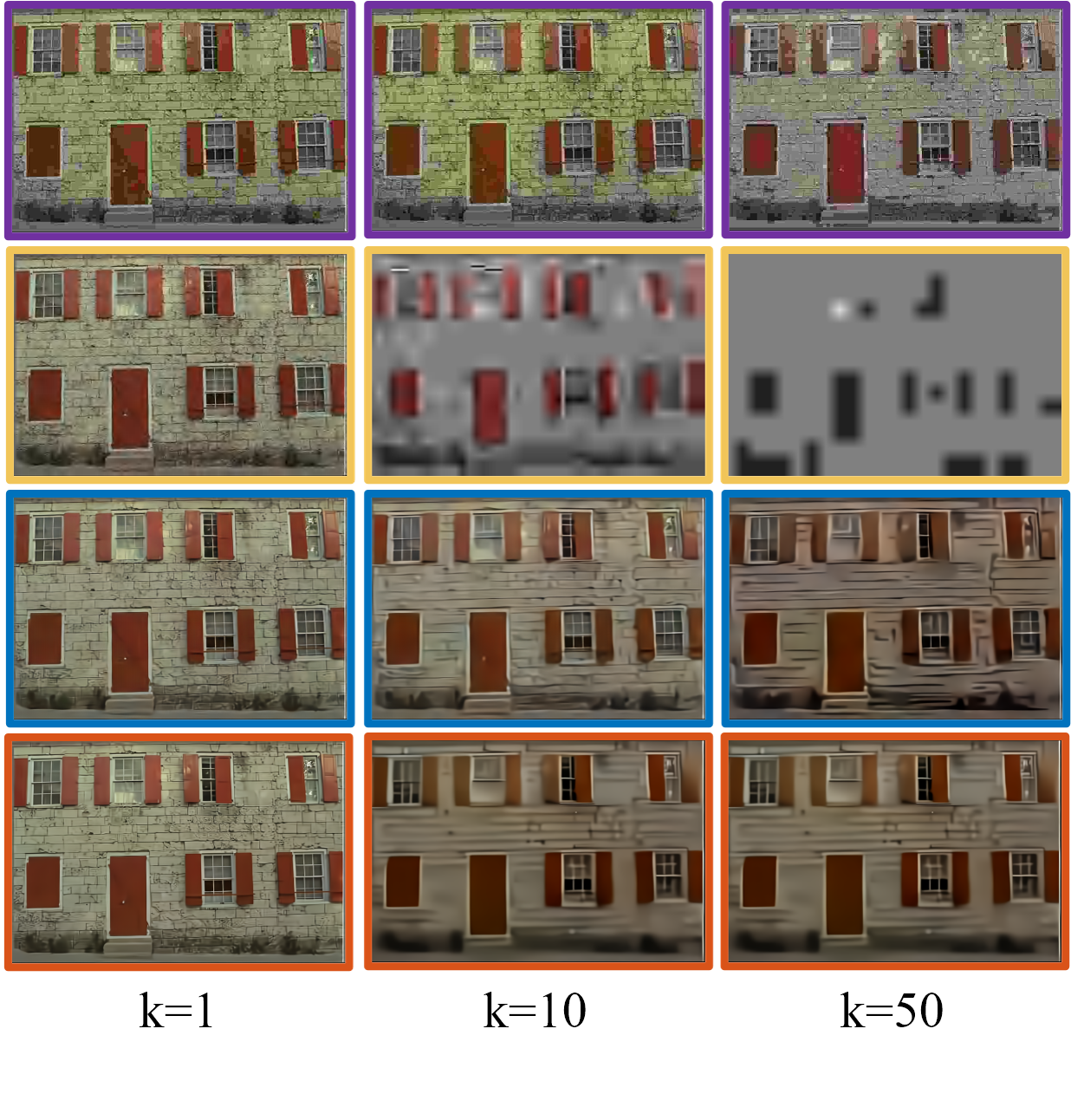}}
\end{minipage}
\caption{(a) The rate-distortion curve obtained by compressing the image multiple times with Cheng20, Ballé18, JPEG2000, and JPEG on the Kodak dataset \cite{kodak1993kodak}. Left: $k=10$. Right: $k=50$. (b) Perceptual results of JPEG, JPEG2000, Ballé18, and Cheng20 models with $k=1,10,50$. $q_{min}$ is selected as the minimal quality possible.}
\end{figure*}

Theorem.~\ref{thm:dist} tells us that the distortion of multi-round compression is always greater than or equal to the distortion of single round compression. And according to the definition of strong idempotence, the equality only holds when the codec satisfies strong idempotence. However, it is important to note that we do not have a bound on the practical bitrate. Although the proof of Theorem.~\ref{thm:dist} tell us the relationship of bitrate lowerbound, the relationship between the actual bitrate is unknown. This means that for non-strong idempotent codec, the bitrate of multi-round compression might be smaller or greater than the bitrate of single round compression. And this further stops us from using the widely applied Bjontegaard metric (BD-Metric) \cite{bjontegaard2001calculation}.

As we need to use the distortion when we describe the strong idempotence, we naturally consider the distortion between the single round compressed image and the multi-round re-compressed image, which is $d(f(\bm{x},q_{min}),f(f(\bm{x},q_1),...,q_k))$. It has several advantages:
\begin{itemize}
    \item \textbf{The metric goes to 0 when strong idempotence is satisfied.} According to the definition of strong idempotence in Eq.~\ref{eq:ide}, the distortion between the single round compressed image and the multi-round re-compressed image is 0. And if we use the distortion between original image and multi-round re-compressed image as metric, we lose this property.
    \item \textbf{The metric upper-bounds the distortion between the original image and multi-round re-compressed image.} Given $d(,.,)$ satisfies triangular inequality, we have 
    \begin{align}
        d(\bm{x},f(\bm{x},q_{min})) &+ d(f(\bm{x},q_{min}),f(f(\bm{x},q_1),...,q_k))  \notag \\ &\ge d(\bm{x},f(f(\bm{x},q_1),...,q_k)),\notag
    \end{align}
    holds, which means that we have an upper bound of the distortion between the original image and multi-round re-compression.
\end{itemize}
Another consideration is that the re-compression quality $q_1,...,q_k$ is stochastic. As we what to know how the codec performs on average, we need the expectation over the metric. Furthermore, the re-compression round $k$ also influences the performance. We are interested in both how a codec performs in short term and long term. And finally, the lowest quality $q_{min}$ also needs to be included. As a codec might perform well in high bitrate but perform not so well in low bitrate.

Taking all above into consideration, we formally define our metric to evaluate strong idempotence of image codec as:
\begin{align}
    \rho(q_{min},k)=\mathbb{E}[d(f(\bm{x},q_{min}),f(f(\bm{x},q_1),...,q_k))], \label{de:def}
\end{align}
where $\rho(q_{min},k)$ is our final metric, and $q_{min}$ is the minimal quality of image and $k$ is the number of re-compression round. Moreover, $q_1,...,q_k$ are uniformly sampled from $\{q_{min},...,q_{max}\}. $As we will show in next section, different $q_{min}$ and $k$ significantly influence the performance. In practice, we approximate the expectation by simple Monte Carlo.
\begin{table*}[t]
\centering
\resizebox{\linewidth}{!}{
\begin{tabular}{@{}lllllllllllllllll@{}}\toprule
&\multicolumn{16}{c}{$\rho(k,q_{min})$ measured in MSE} \\ \cmidrule{2-17}
 & \multicolumn{8}{c}{$k=10$} & \multicolumn{8}{c}{$k=50$} \\ \cmidrule(rr){2-9} \cmidrule(ll){10-17}
 $q_{min}=$ & 1 & 2 & 3 & 4 & 5 & 6 & 7 & 8  & 1 & 2 & 3 & 4 & 5 & 6 & 7 & 8 \\ \midrule
\textit{Traditional} & &   &   &   &   &   &   & & &   &   &   &   &   &   &  \\
JPEG     &288.2 &137.7 &136.4 &96.3 &62.9 &51.4 &70.6 &39.5 &552.9 &354.7 &331.0 &283.1 &187.4 &216.7 &156.0 &97.2\\
JPEG2000 &1308 &669.7 &455.1 &222.5 &172.5 &120.5 &90.2 &69.7 &3098 &3044 &2396 &1811 &823.2 &471.6 &315.1 &226.3  \\ \midrule
\textit{Learning-based} & &   &   &   &   &   &   & & &   &   &   &   &   &   &  \\
Ballé18 \cite{balle2018variational}  &108.7 &81.2 &84.8 &74.5 &61.9 &46.4 &37.8 &48.2 &877.4 &852.7 &798.5 &808.4 &816.8 &719.0 &848.9 &1342\\
Cheng20 \cite{cheng2020learned} &123.6 &109.3   &95.9   &83.4   &62.9   &41.2  & - & - &1030  &900.5 &835.7 &789.7 &676.6 &682.9 & - & -\\ \bottomrule
\end{tabular}
}
\caption{The value of $\rho(q_{min},k)$ measured in MSE using k equal to 10 and 50 for Cheng20, Ballé18, JPEG2000, and JPEG on the Kodak dataset \cite{kodak1993kodak}.}
\label{tab:01}
\end{table*}
\subsection{Testing Protocol}
We specify a testing protocol to guide the evaluation of strong idempotence of an image codec. Specifically, we give the following procedure:
\begin{itemize}
    \item First, prepare the testing dataset. We recommend the Kodak dataset \cite{kodak1993kodak} to be used. 
    \item Second, select list of $q_{min}$ and $k$ to evaluate each $\rho(q_{min},k)$. We recommend to uniformly sample $8$ $q_{min}$ from the quality points of codec. And for $k$, we recommend $k=10$ and $k=50$ for short and long term performance.
    \item Third, uniformly sample $k$ quality $q_i$ from range $\{q_{min},q_{max}\}$. Compress the image once with $q_{min}$. Then, compress the image repeatly according to $q_i$s and evaluate the distortion between once compression and repeat compression $\hat{\rho(q_{min},k)}$. Repeat this process for $b$ times and take the average of distortion to get the final $\rho(q_{min},k)$.
    \item Fourth, repeat step Third for all selected $q_{min},k$ to complete evaluation.
\end{itemize}

\section{Experiments}
\subsection{Experiment Setup}
Following the above testing protocol, we evaluate the strong idempotence of two traditional codecs and two LIC. The two traditional codecs are JPEG and JPEG2000. We use libjpeg to evaluate JPEG comes from , and openjpeg1.5.1. to evaluate JPEG2000. The LIC models are Ballé18 \cite{balle2018variational} with mean scale hyperprior and Cheng20 \cite{cheng2020learned} and with pre-trained model by \cite{begaint2020compressai}. Those two models are widely adopted as baseline in LIC community. 

For JPEG and JPEG2000, the lowest quality is uniformly sampled in the interval with bpp (bits-per-pixel) range from $0$ to $3.5$. For JPEG, we test 8 $q_{min}=\{5, 15, 25, 35...75\}$. For JPEG2000, we test 8 $q_{min}=\{24, 26, 28, 30...38\}$. For Ballé18 \cite{balle2018variational} and Cheng20 \cite{cheng2020learned}, the minimal quality level $q_{min}$ traverses all model quality, which are $q_{min}=\{1,...,6\}$ and $\{1,...,8\}$ respectively. We set parameter k (the re-compression times), as 10 and 50. For the Monte Carlo estimation of each $\rho(q_{min},k)$, we sample $b=10$ times for JPEG2000, Ballé18 \cite{balle2018variational} and Cheng20 \cite{cheng2020learned}. However, we sample 50 times for JPEG due to its high variance. (See Fig.~\ref{fig:02a}) 

\subsection{Experiment Results}
The evaluation results are presented in Fig.~\ref{fig:02a}, Fig.~\ref{fig:02b} and Tab.~\ref{tab:01}. Fig.~\ref{fig:02a} compares the rate-distortion curves of direct compress with $q_{min}$ and multi-round compression with minimal quality $q_{min}$. 

The re-compression R-D curve of JPEG2000 is very close to the single compression. This means that the overall R-D performance of JPEG2000 is only marginally harmed by multi-round compression. However, it is also notable that the absolute quality decay of JPEG2000 is also very fast. For the highest $q_{min}$ and $k=10$, the PSNR of JPEG2000 drops more than $5$ dB. On the other hand, JPEG \cite{wallace1992jpeg} also suffers from significant quality decay, while its bitrate remains unchanged. Compared with traditional codecs, the LIC methods are more sensitive to multi-round compression. Both Ballé18 \cite{balle2018variational} and Cheng20 \cite{cheng2020learned} suffers from very significant quality decay when $k=10$. And for $k=50$, the R-D curves of  Ballé18 and Cheng20 are no longer monotonously increasing, which means that when the bitrate goes higher, the quality goes lower. 

Table 1 shows the value of $\rho(q_{min},k)$ measured in MSE. Compared with $\rho(q_{min},k=10)$s, the $\rho(q_{min},k=50)$s are significantly lower. There are also slight fluctuations in the last few points. Fig.~\ref{fig:02b} shows the perceptual results of multi-round compression with the lowest quality level $q_{min}$in the test interval with $k=1,10,50$. As $k$ increases, the perceptual quality of traditional codecs falls faster than LIC, which does not conform the subjective R-D curve. This phenomenon is particularly evident for JPEG2000. As we can see, the image is barely recognizable for JPEG2000 after $k=50$ re-compression.

\section{Conclusion}
In this paper, we propose the concept of strong idempotent codec and study its properties. We propose a metric to evaluate strong idempotence of image codec and formalize a testing protocol. Finally, we evaluate the strong idempotence of current image codecs, including two traditional codecs and two recent neural codecs. Experimental results show that the current codecs are very sensitive to multi-round re-compression with different qualities. And we encourage future works to improve this condition.
\bibliographystyle{IEEEbib}
\bibliography{strings,refs}

\end{document}